\documentclass[AMA,STIX1COL]{WileyNJD-v2}
\usepackage{graphicx}
\usepackage{rotating,stackengine,scalerel}
\usepackage{caption}
\usepackage{soul}
\usepackage{booktabs}
\usepackage{float}

\usepackage{draftwatermark}
\SetWatermarkText{DRAFT}
\SetWatermarkScale{1}
\SetWatermarkLightness{1}

\newcommand{\ffrac}[2]{\ensuremath{\frac{\displaystyle #1}{\displaystyle #2}}}
\newcommand\wye{\scalerel*{\stackengine{-1pt}{%
  \rotatebox[origin=c]{30}{\rule{10pt}{.9pt}}\kern-1pt%
  \rotatebox[origin=c]{-30}{\rule{10pt}{1.3pt}}}{%
  \rule{.9pt}{10pt}}{O}{c}{F}{F}{S}}{\Delta}}
 
\articletype{Research Article}%

\received{26 April 2016}
\revised{6 June 2016}
\accepted{6 June 2016}

\begin{document}

\title{Weighted trajectory analysis and application to clinical outcome assessment}


\author[1]{Utkarsh Chauhan}

\author[2]{Kaiqiong Zhao}

\author[3]{John Walker}

\author[3]{John R. Mackey*}

\authormark{CHAUHAN \textsc{et al}}

\address[1]{\orgdiv{Faculty of Medicine and Dentistry}, \orgname{University of Alberta}, \orgaddress{\state{Edmonton}, \country{Canada}}}

\address[2]{\orgdiv{Department of Mathematical and Statistical Sciences}, \orgname{University of Alberta}, \orgaddress{\state{Edmonton}, \country{Canada}}}

\address[3]{\orgdiv{Division of Medical Oncology}, \orgname{University of Alberta}, \orgaddress{\state{Edmonton}, \country{Canada}}}

\corres{*John R. Mackey, Division of Medical Oncology, University of Alberta, Cross Cancer Institute, 11560 University Ave NW, Edmonton, AB, T6G 1Z2. \email{john.mackey@albertahealthservices.ca}}


\abstract[Summary]{The Kaplan-Meier estimator (KM) is widely used in medical research to estimate the survival function from lifetime data. KM is a powerful tool to evaluate clinical trials due to simple computational requirements, a logrank hypothesis test, and the ability to censor patients. However, KM has several constraints and fails to generalize to ordinal variables of clinical interest such as toxicity and ECOG performance. We devised Weighted Trajectory Analysis (WTA) to combine the advantages of KM with the ability to visualize and compare treatment groups for ordinal variables and fluctuating outcomes. To assess statistical significance, we developed a new hypothesis test analogous to the logrank test. We demonstrate the functionality of WTA through 1000-fold clinical trial simulations of unique stochastic models of chemotherapy toxicity and schizophrenia disease course. At increments of sample size and hazard ratio, we compare the performance of WTA to KM and Generalized Estimating Equations (GEE). WTA generally required half the sample size to achieve comparable power to KM; advantages over GEE include its robust non-parametric approach and summary plot. We also apply WTA to real clinical data: the toxicity outcomes of melanoma patients receiving immunotherapy and the disease progression of patients with metastatic breast cancer receiving ramucirumab. The application of WTA demonstrates that using traditional methods such as KM can lead to both Type I and II errors by failing to model illness trajectory. This article outlines a novel method for clinical outcome assessment that extends the advantages of Kaplan-Meier estimates to ordinal outcome variables.}


\maketitle


\section{Introduction}\label{sec1}

The Kaplan-Meier estimator (KM)\cite{kaplanmeier}, also referred to as the product-limit estimator, is widely used in medical research to estimate the survival function from lifetime data. KM is a non-parametric approach for time-to-event data that are often not normally distributed. To generate the KM estimates, time-to-event and the status of each subject at the last observed timepoint are needed.\cite{peto1} The event of interest may be death from any cause when we are determining overall survival, and death due to specific cause for cause-specific survival. KM estimates are frequently used in oncology and other medical disciplines. KM is used to compare two or more treatment arms in clinical trials using the logrank test.\cite{peto2} Patients that exit the trial without having experienced the event of interest at last follow-up are censored and omitted from further estimates.

The relatively simple computational requirements for KM provide a powerful method to estimate time-to-event data. However, the advantages of KM in clinical research cannot be extended to important ordinal outcomes such as toxicity grade and Eastern Cooperative Oncology Group (ECOG) performance status.\cite{ecog} Ordinal outcome variables are ubiquitous in medicine to measure patient health status over time, but no statistical methods exist that combine censoring, graphical comparison of trajectories, and hypothesis testing for these variables. Often, ordinal clinical outcomes are collapsed to binary definitions to facilitate the use of KM; this causes information loss, introduces an arbitrary cutpoint, and may lead to inaccurate conclusions. New methods are required to map the trajectory of ordinal outcomes and compare treatment arms in clinical trials.

The KM method has three conditions that limit its generalizability to other variables of interest in clinical research:

\begin{enumerate}

	\item \textbf{Binary Condition}

The event must be binary in nature or coded into binary form (0 for non-occurrence, 1 for occurrence). It is not possible to capture grades or stages of severity. For example, death is naturally binary (0 for alive, 1 for dead), but an outcome variable such as toxicity (measured in grades from 0 to 4) must be coded into binary form by setting a threshold for event occurrence, such as arbitrarily defining an event as any toxicity exceeding grade 2.

	\item \textbf{Descent Condition}

Event occurrence always produces a drop in the KM curve (a consequence of plotting probability). It is not possible to track the trajectory of conditions that can both improve and worsen over time. For example, patients experiencing rising toxicity due to chemotherapy require additional interventions to tolerate therapy. The interventions may initially improve symptoms and reduce toxicity grade, but fail to sustain benefits in subsequent treatment cycles. In the KM estimate following the above example, this complex trajectory is simplified to an event occurrence the first time toxicity increases beyond grade 2.

	\item \textbf{Finality Condition}

Once a patient experiences the event of interest, they are omitted from any subsequent analysis.

\end{enumerate}

Weighted Trajectory Analysis (WTA) is a method that combines the simplicity and practicality of KM with the ability to compare treatment groups for ordinal variables and bidirectional outcomes. Trajectories are presented using plots that track health status for treatment arms over time. WTA permits the censoring of patients that exit the study. To determine statistical significance, we developed a “weighted” logrank test.

In section 2, we describe the methodology and theory of KM and WTA along with their respective hypothesis tests and provide a computational approach to WTA robust to smaller datasets. We also outline GEE longitudinal analysis prior to its use as an additional comparator to WTA in subsequent simulation studies. In sections 3 and 4, we describe unique simulation studies with chemotherapy toxicity grade and schizophrenia symptom stage as the variables of interest, respectively. In section 5, we apply WTA to real clinical datasets: first with the toxicity outcomes of melanoma patients receiving different immunotherapy protocols, and second with tumour response outcomes of patients with metastatic breast cancer receiving an anti-angiogenic drug. Finally, we discuss the results and implications of both our simulations and real-world analyses in Section 6.

\section{Methodology and Theory}\label{sec2}

\subsection{Kaplan-Meier Estimator}

The goal of the Kaplan-Meier estimator (KM) is to estimate a population survival curve from a sample with incomplete time-to-event observations.\cite{kaplanmeier}  “Survival” times need not relate to death, but can refer to any event of interest, such as local recurrence or stroke. The event in this instance is a binary variable, meaning that samples have either experienced the event up to a given time point or not. The times to failure on each subject are thus characterized by two variables: (1) serial time, (2) outcome of event occurrence or censorship.

Suppose that  $t_{0} < t_{1} < t_{2} < ... < t_{K}$ be the $K$ distinct failure times observed in the sample. We write $n_{j}$ and $d_{j}$ as the number of patients at risk and number of events at time $t_{j}$, respectively, where $j = 1, 2, ..., K$. Note that the patients who are lost to follow-up or withdraw from the trial before experiencing the event of interest (i.e. censored samples) are taken out of the risk set at the subsequent time points. 

The KM estimate at time $t_{j}$, $\widehat{S(t_j)}$, is calculated as the cumulative survival probability up to and including time $t_{j}$,
\\
\begin{equation}
S(t_{k}) = \prod_{j=1}^{k} \left (1-\frac{d_{j}}{n_{j}}  \right ),
\label{itersurvival}
\end{equation}
\\
where $S(t) = 1$ for $t < t_{1}$. The Kaplan-Meier curve is plotted as a stepwise function representing the change in survival probability over time. 

To compare treatment arms, multiple survival functions are plotted together, enabling the comparison of differences in survival experience between groups. Treatment options can be compared using metrics such as median survival and hazard ratios. The logrank test is used to assess if the differences are statistically significant: this test and its modification for WTA are discussed in Sections 2.4 and 2.5, respectively.

\subsection{Weighted Trajectory Analysis}

Weighted Trajectory Analysis (WTA) is a modification of KM that provides the following advantages:

\begin{itemize}
\item Assesses outcomes defined by various ordinal grades (or stages) of clinical severity;
\item Permits continued analysis of participants following changes in the variable of interest;
\item Demonstrates the ability of an intervention to both prevent the exacerbation of outcomes and improve recovery, and the time course of these effects.
\end{itemize}

Several properties of KM curves crucial for clinical trial evaluation are incorporated within WTA. The test is non-parametric and provides the ability to censor patients that withdraw or are lost to follow-up. Outcomes for various treatment arms can be assessed using a summary plot that depicts all patients in serial time. The test for significance is a modification of the logrank test by Peto et al., which is the standard method for comparing KM survival curves.\cite{peto2} The logrank test is described in Section 2.4, and the weighted logrank test follows in Section 2.5. As the analytical form of the test is a conservative estimate that operates under the normal approximation, a more computationally intensive simulated approach is outlined in Section 2.6.

In WTA, an event is a change in grade or stage, or more generally, a severity score. The severity score must be ordinal but can have an arbitrary range of severity that depends on the variable of interest (for example, I-IV for heart failure class\cite{nyha}). Unlike KM, an event does not omit the patient from subsequent analysis. Both increases and decreases in variables of interest are captured as events. Participants can enter trajectory analysis at any starting stage, though inferences on trial results are most powerful if treatment arms are randomized to the same median starting stage.

Redefining the event allows clinical assessment of the overall trajectory of a group of patients, mapping both deterioration and improvement in health status over time. Graphically, the staircase representing survival in the Kaplan-Meier estimator always descends. The WTA staircase can both descend and rise over time to capture the dynamics of a patient’s clinical status.

Variables of interest include any ordinal outcome variables with a defined, finite range. Examples include ECOG Performance\cite{ecog} and Common Terminology Criteria for Adverse Events (CTCAE) toxicity scores,\cite{ctcae} both with ordinal scoring that ranges from 0 to 5.

For this reason, a binary variable such as death (0, alive vs. 1, dead) is not an appropriate variable of interest. In this circumstance, the range of the ordinal variable is set to 1, and the modified significance test reduces to the standard logrank test. Conversely, ECOG performance is an appropriate variable of interest, given that it is ordinal with a defined range and can both improve and worsen over time. In WTA, a higher score in the variable of interest generally represents poorer health status. Variables that follow the opposite trend can be adapted to WTA by simply reversing the polarity of the ordinal scale.

Censoring in WTA is similar to KM. Patient loss to follow-up and withdrawal requires censoring, but patients may experience several events prior to being censored. Censoring is represented on the plot using a Wye-symbol ($\wye$). The number of patients remaining within the study is tabulated below the plot at evenly spaced time intervals for each treatment arm.

\subsection{Mathematical Overview of Weighted Trajectory Analysis}

Weighted Trajectory Analysis plots the health status of treatment arms as a function of time. Time values must be discrete, but can correspond to days, weeks, months, or any chosen interval. For each time value on the x-axis, there is a corresponding score on the y-axis: a weighted health status. The higher the weighted health status, the healthier the group. This score is scaled by the initial size of the treatment arm to facilitate simple comparison of groups with unequal size.

Consider a group of $n$ patients with toxicity grade ranging from Grade 0 (asymptomatic/mild toxicity) to Stage 5 (death related to an adverse event). The weighted health status at time point $j$ is denoted by $U_{j}$, where j = 0, 1, ..., $z$. For each treatment arm, $U_{j}$ has a maximum value of 1 and a minimum value of 0. Suppose we begin a trial with all patients absent disease burden at Grade 0: $U_{j}$ = $U_{0}$ = 1. A trial with the highest possible morbidity requires all patients to experience a Grade 5 toxicity (death): at this point, $U_{j}$ will drop to 0.

We let $g_{i,j}$ represent the severity score for the $i$th patient at time $j$, i = 1, ..., $n$. The severity score is identical to their ordinal score for the variable of interest. If the range of the ordinal variable of interest does not have 0 as one extreme end, all values must be shifted to set 0 as the starting score (the polarity may also be reversed so that 0 represents peak health status). All patients begin the trial at Grade 0, which reflects $g_{i,0}$ = 0. If a patient labeled with index 50 has a Grade 3 injury on the seventh time point, their severity score $g_{50,7}$ = 3. 

Scaling for the WTA curve is performed through normalizing to a minimum of 0 and a maximum of 1 by using the initial weight of the treatment arm. This weight, $w_{0}$, is the product of starting patient count $n_{0}$ and the range of the ordinal variable of interest $r$:
\\
\begin{equation}
w_{0} = n_{0}r.
\label{w0}
\end{equation}
\\
Suppose the initial size of the group, $n_{0}$, is 100 patients. The range $r$ in the ordinal variable (toxicity grade) is 5. Then, $w_{0}$ is 500. The value of the weight changes over time due to patient censoring reflected by a drop in $n_{j}$. The general equation for $w_{j}$ is provided in Section 2.5 and is used in the weighted logrank test. However, for scaling and plotting $U$, only the initial weight of a given treatment arm, $w_{0}$, is required.

The initial value $U_{0}$ is a perfect score of 1.

\begin{equation}
U_{0} = 1
\label{u0}
\end{equation}
\\
Subsequent values of $U$ deviate based on observed event occurrences $d_{j}$. We define event occurrence as a change in the variable of interest for a given patient $i$ at time $j$:
\\
\begin{equation}
d_{i,j} = g_{i,j+1} - g_{i,j}.
\label{dj}
\end{equation}
\\
Therefore, the observed event score for a group of $n$ patients is defined as
\\
\begin{equation}
d_{j} = \sum_{i=1}^{n}d_{i,j} = \sum_{i=1}^{n}(g_{i,j+1} - g_{i,j}),
\label{dj}
\end{equation}
\\
with patients censored following time $j$ not contributing to the sum. Events and resulting changes in treatment arm trajectory are always scaled by $w_{0}$. Using this event definition, $U_{j}$ can be calculated iteratively from $U_{0}$:
\\
\begin{equation}
U_{j+1} = U_{j} - \frac{d_{j}}{w_{0}}, j = 0, 1, 2...
\label{ujiter}
\end{equation}
\\
Alternatively, $U_{j}$ for any given time point can be computed as follows:
\\
\begin{equation}
U_{j} = 1 - \ffrac{\sum_{j=0}^{j-1}d_{j}}{w_{0}}, j \epsilon \mathbb{Z}^{+}.
\label{uj}
\end{equation}
\\
Values for $d_{j}$ at a given time point can be negative, and these represent cases in which the treatment arm improved in overall health status. From Equations \ref{ujiter} and \ref{uj} it follows that a negative value of $d_{j}$ produces an increase in the weighted health status $U_{j}$.

\subsection{The Logrank Test}

We present here the standard formula of the log-rank test statistic.

\begin{itemize}
\item Let  $t_1<t_2<{\dots}<t_K$ be $K$ distinct failure times observed in the data
\item  $n_j^A$ is the number of patients in group A at risk at  $t_j$, where  $j=1,2,{\dots},\mathit{K.}$
\item  $n_j^B$ is the number of patients in group B at risk at  $t_j$, where  $j=1,2,{\dots},\mathit{K.}$
\item  $n_j=n_j^A+n_j^B$ is the total number of patients at risk at  $t_j$, where $j=1,2,{\dots},\mathit{K.}$
\item  $d_j^A$ is the number of patients who experienced the (binary) event in group A at $t_j$.
\item  $d_j^B$ is the number of patients who experienced the (binary) event in group B at $t_j$.
\item  $d_j=d_j^A+d_j^B$ is the total number of patients who experienced the (binary) event at $t_j$.
\item  $S^A(t)$ and  $S^B(t)$ are the survival functions for group A and B, respectively.
\end{itemize}

The information at $t_{j}$ can be summarized in a 2×2 table.

\begin{table}[h]
\centering
\begin{tabular}{@{}|c|c|c|c|@{}}
        & Observed to fail at $t_j$ &                         & At risk at $t_{j}$ \\
Group A & $d_{j}^{A}$               & $n_{j}^{A} - d_{j}^{A}$ & $n_{j}^{A}$        \\
Group B & $d_{j}^{B}$               & $n_{j}^{B} - d_{j}^{B}$ & $n_{j}^{B}$        \\
        & $d_{j}$                   & $n_{j} - d_{j}$         & $n_{j}$           
\end{tabular}
\end{table}

Under the null hypothesis  $H_0:S^A(t)=S^B(t),d_j^A$ \ follows a \textit{hypergeometric distribution} conditional on the margins  $(n_j^A,n_j^B,d_j,n_j-d_j)$. The expectation and variance of  $d_j^A$ take the form

\begin{equation}
e_j^A=E\left(d_j^A\right)=n_j^A\frac{d_j}{n_j}
\end{equation}
\begin{equation}
V_j=\mathit{Var}\left(d_j^A\right)=\frac{n_j^An_j^B(n_j-d_j)}{n_j^2(n_j-1)}d_j.
\end{equation}
\\
Define the observed aggregated number of failures in group A as

\begin{equation}
O^A=\sum _{j=1}^Kd_j^A.
\end{equation}
\\
The expected aggregated number of failures in group A is thus

\begin{equation}
E\left(O^A\right)=E^A=\sum _{j=1}^Ke_j^A.
\end{equation}
\\
The contributions from each  $t_j$ are independent and thus the variance of  $O^A$ is

\begin{equation}
\mathit{Var}\left(O^A\right)=V=\sum _{j=1}^KV_j.
\end{equation}
\\
Under the null hypothesis  $H_0:S^A\left(t\right)=S^B\left(t\right),$ \ the log-rank test statistic

\begin{equation}
Z=\frac{O^A-E^A}{\sqrt V}=\frac{\sum _{j=1}^K(d_j^A-e_j^A)}{\sqrt{\sum _{j=1}^KV_j}}{\sim}N\left(0,1\right).
\end{equation}
This is an asymptotic result derived from the central limit theorem (CLT). Note that replacing  $O^A$ and $E^A$ with 
$O^B$ and $E^B$ leads to the exact same p-value.

The extension to ordinal event in the following section is based on this Z test statistic. 

\subsection{The Weighted Logrank Test - Analytical Method}

We define an event as a change in the severity score of a given condition. Let  $g_{i,j}^A$ \ be the severity score for
the  $i^{\mathit{th}}$ \ individual in group A at time  $t_j$, where  $i=1,2,{\dots},n_j^A$ \ and  $j=1,2,{\dots},K$.
\ Define  $d_{i,j}^A$ \ as the change in the severity score from time  $t_{j+1}$ \ to  $t_j$. 

\begin{equation}
d_{i,j}^A=g_{i,j+1}^A-g_{i,j}^A, j=1,2,K-1.
\end{equation}
\\
Without loss of generality, we consider a severity score ranging from Stage 0 to Stage 4. As a result,  $d_{i,j}^A$
\ has a total of 9 possible values  $-4,-3,-2,-1,0,1,2,3,4$, if the observation of this person is uncensored at 
$t_{j+1}.$

\begin{itemize}
\item Let  $L$ \ be the total number of possible values taken by the change variable  $d_{i,j}^A$. When a severity score
takes values from 0 to 4,  $L=9$. 
\item Let  $W$ \ be the ordered non-decreasing list of the  $L$ \ possible change values. When a severity score takes
values from 0 to 4,  $W=(-4,-3,-2,-1,0,1,2,3,4)$. 
\item Let  $w_l$ \ be the  $l^{\mathit{th}}$ \ element of  $\mathit{W.}$
\item Let  $d_j^{A,l}$ \ be the number of subjects in group A at  $t_j$ \ whose change values equal to  $w_l$:
\begin{equation}
d_j^{A,l}=\sum _{i=1}^{n_j^A}d_{i,j}^AI(d_{i,j}^A=w_l)
\end{equation}
\\
where,  $I\left(d_{i,j}^A=w_l\right)=1$, when $d_{i,j}^A=w_l$ and equal to 0, otherwise. 
\item Let  $d_j^{B,l}$ \ be the number of subjects in group B at  $t_j$ \ whose change values equal to  $w_l$.
\item  $d_j^{\left(l\right)}=d_j^{A,l}+d_j^{B,l}$ is the total number of patients whose change values equal to  $w_l$
\ at  $t_j$.
\end{itemize}

The information at  $t_j,j=1,2,{\dots},K-1$ can be summarized in a 2 x 10 table:

\begin{table}[h]
\centering
\begin{tabular}{@{}|c|c|c|c|c|c|c|c|c|c|c|c|@{}}
\toprule
Observed values of $d_{i,j}$ ($w_l$) &
  -4 &
  -3 &
  -2 &
  -1 &
  0 &
  1 &
  2 &
  3 &
  4 &
   &
  At risk at $t_j$ \\ \midrule
Group A &
  $d_{j}^{A,1}$ &
  $d_{j}^{A,2}$ &
  $d_{j}^{A,3}$ &
  $d_{j}^{A,4}$ &
  $d_{j}^{A,5}$ &
  $d_{j}^{A,6}$ &
  $d_{j}^{A,7}$ &
  $d_{j}^{A,8}$ &
  $d_{j}^{A,9}$ &
  $n_j^A-\sum _{l=1}^Ld_j^{A,l}$ &
  $n^A_j$ \\ \midrule
Group B &
  $d_{j}^{B,1}$ &
  $d_{j}^{B,2}$ &
  $d_{j}^{B,3}$ &
  $d_{j}^{B,4}$ &
  $d_{j}^{B,5}$ &
  $d_{j}^{B,6}$ &
  $d_{j}^{B,7}$ &
  $d_{j}^{B,8}$ &
  $d_{j}^{B,9}$ &
  $n_j^B-\sum _{l=1}^Ld_j^{B,l}$ &
  $n^B_j$ \\ \midrule
 &
  $d_j^{(1)}$ &
  $d_j^{(2)}$ &
  $d_j^{(3)}$ &
  $d_j^{(4)}$ &
  $d_j^{(5)}$ &
  $d_j^{(6)}$ &
  $d_j^{(7)}$ &
  $d_j^{(8)}$ &
  $d_j^{(9)}$ &
  $n_j-\sum _{l=1}^Ld_j^{(l)}$ &
  $n_j$ \\ \bottomrule
\end{tabular}
\end{table}

Under the null hypothesis  $H_0:S^A\left(t\right)=S^B\left(t\right)$, $\left(d_j^{A,1},d_j^{A,2},d_j^{A,3},{\dots},d_j^{A,L}\right)$ follows a \textit{multivariate hypergeometric distribution} conditional on the margins $\left(n_j^A,n_j^B,\left\{d_j^{\left(l\right)}\right\}_{l=1}^L,n_j-\sum
_ld_j^{\left(l\right)}\right)$.

\bigskip

We can show that the mean and variance of  $d_j^{A,l}$, where $l{\in}\left\{1,2,{\dots},L\right\}$, are

\begin{equation}
e_j^{A,l}\ {\triangleq}\ E\left(d_j^{A,l}\right)=n_j^A\frac{d_j^{(l)}}{n_j}
\end{equation}
\\
\begin{equation}
\sigma _{j,ll}\ {\triangleq}\ \mathit{Var}\left(d_j^{A,l}\right)=\frac{n_j^An_j^B(n_j-d_j^{(l)})}{n_j^2(n_j-1)}d_j^{(l)}.
\end{equation}
\\

For distinct  $l,q{\in}\left\{1,2,{\dots},L\right\}$, we can derive the covariance of $d_j^{A,l}$ and  $d_j^{A,q}$

\begin{equation}
\sigma_{j,lq}\ {\triangleq}\ \mathit{Cov}\left(d_j^{A,l},d_j^{A,q}\right)=-\frac{n_j^An_j^B}{n_j^2(n_j-1)}d_j^{\left(l\right)}d_j^{\left(q\right)},\ l{\neq}q.
\end{equation}
\\
These moment results are derived from the definition of multivariate hypergeometric distribution. To account for the direction and the magnitude of the change variable, we define the \textit{observed} \textbf{weighted changes} as

\begin{equation}
O_j^w=\sum _{l=1}^Lw_ld_j^{A,l}.
\end{equation}
\\
When a severity score is defined as a range from 0 to 4, the weights  $w_l$ takes the values of $(-4,-3,-2,-1,0,1,2,3,4)$ for  $l=1,2,{\dots},9.$. The expected value of  $O_j$ can be written as

\begin{equation}
E_j^w=\sum _{l=1}^Lw_le_j^{A,l}.
\end{equation}
\\
When the event is coded as a binary outcome, this weighted changes $O_j^w$ is reduced to the $e_j^A$ defined above. Using the results in (1) and (2), we can write the variance of the weighted score  $O_j^w$ \ as

\begin{equation}
V_j^w=\mathit{Var}\left(O_j^w\right)=\sum _{l=1}^L\sum _{q=1}^Lw_lw_q\sigma _{j,\mathit{lq}},
\end{equation}
\\
where,  $\sigma _{j,\mathit{lq}}$ \ is defined in equation (2) when  $l{\neq}q$ and is
defined in equation (1) when  $l=q$.

Similarly, we can aggregate the observed/expected weighted changes across all  $K$ time points and define a $Z$ test statistic. The weighted log-rank test statistic is defined as 

\begin{equation}
Z=\frac{\sum _{j=1}^K\left(O_j^w-E_j^w\right)}{\sqrt{\sum _{j=1}^KV_j^w}},
\end{equation}
\\

which follows the standard normal distribution $N\left(0,1\right),$ under the null hypothesis 
$H_0:S^A\left(t\right)=S^B\left(t\right).$ Equivalently,

\begin{equation}
Z^2=\frac{\left[\sum _{j=1}^K\left(O_j^w-E_j^w\right)\right]^2}{\sum _{j=1}^KV_j^w}{\sim}\chi _1^2,
\label{t_stat}
\end{equation}
\\
i.e. the square of the Z test statistic follows a chi-square distribution with 1 degrees of freedom.

The asymptotic result in equation (3) is based on the assumption that the total number of distinct failure times recorded in the pooled samples (i.e. $K$) is sufficiently large. For smaller trials with shorter follow-up periods, this analytical method can provide conservative conclusions and result in Type II errors below the designated significance level, as demonstrated in Section 3.3. To complement the analytical method, we also propose a bootstrap-based approach for calculating p-values, which, despite requiring more computational effort, remains accurate and sensitive independent of trial sizes.

\subsection{The Weighted Logrank Test - Computational Method}

A completed trial can be analyzed either instantly by the analytical approach or through rigorous simulations in a more sensitive computational approach. Compared to the design phase, the advantage of a completed trial is the wealth of collected data. Multistate Markov modelling (MSM), available in the \textit{msm} package in R, provides a powerful method to compute transition intensities of an inputted dataset through maximum likelihood estimation. The steps to analyze a complete trial are as follows:

\begin{enumerate}
\item Determine transition probabilities using \textit{msm} to load into n-fold simulations blind to treatment assignment
\item Generate a distribution of the null hypothesis using the test statistic, equation (\ref{t_stat})
\item Calculate a test statistic from the clinical data and then determine a p-value by comparison to distribution of the null
\end{enumerate}

Software with build-in tools to facilitate analytical and computational methods to streamline the use of WTA for investigators is in production.

\subsection{GEE Longitudinal Analysis}

Generalized Estimating Equation (GEE) (Liang and Zegar 1986) has been a widely used regression-based tool for analyzing longitudinal data.\cite{gee} We compare the performance of our weighted trajectory approach to GEE method. In GEE, we model the severity score as outcomes and treatment group as covariate. We specify the autoregressive correlation structure to account for the dependence among the severity measures from the same patient. We use an identity mean-variance link function and leave the scale parameter unspecified.  The significance test for the association between patients’ severity score and treatment status is carried out using a Wald test statistic with the sandwich variance estimator.

A major advantage of GEE over likelihood-based methods (e.g., multi-state models), a is that the joint distribution of longitudinal outcomes does not have to be fully specified. Therefore, if the mean structure is accurately specified, the mean parameters, e.g. the treatment effect in our case, can be consistently estimated, regardless of whether or not the covariance structure is correctly characterized. Our weighted log-rank test is more robust than GEE, because it is a nonparametric test and does not make any assumptions about the survival outcomes. In addition, a visual representation of the survival trajectory over time is naturally accompanied by our proposed test statistic, which tracks of the number of changes in the severity score over time. On the other hand, GEE enables simultaneous modeling of multiple covariates, while our approach focuses on comparison between two treatment groups. In the following simulation studies, we make a direct performance comparison between GEE and WTA.

\section{Simulation Study 1 - Toxicity}\label{sec3}

In our first clinical trial simulation study, we demonstrate the functionality of WTA and present its advantages over KM analysis. We establish the strength of our novel method through rigorous power comparison between KM, GEE, and both analytical and simulated approaches to WTA.

The design is a Phase III comparison of toxicity outcomes from chemotherapy between two treatment arms (control and treatment, 1:1 allocation). The variable of interest is CTCAE toxicity: grades range from 1 (mild/no toxicity) to 5 (death from toxicity).\cite{ctcae} For example, grades of oral mucositis range from (1) asymptomatic/mild, (2) moderate pain or ulcer that does not interfere with oral intake, (3) severe pain interfering with oral intake, (4) life threatening consequences indicating urgent intervention, and (5) death. For the purposes of WTA, the ordinal range of 1-5 is shifted to 0-4, and thus censoring takes place at grade 4.

The simulation study was generated using Python 3.7.\cite{python} Study simulations are a stochastic process in which randomly generated numbers are programmed to mirror fluctuating toxicities experienced by groups of patients undergoing chemotherapy cycles with daily measurements of treatment toxicity. Each instance of the simulation requires a specified hazard ratio and sample size prior to the stochastic generation of toxicity. Table 2 provides a snapshot of the results for a single simulated clinical trial.

Each patient (represented by an ID number) has a risk of developing treatment toxicity over time. This risk is determined by their treatment group and the numbers of days they have spent in the study. The values within Table 2 were assigned as follows:

\begin{enumerate}
\item Treatment group: randomly assigned as 0 or 1 with the constraint of having an equal number of patients allocated to each group.
\item Duration: the number of days a patient remains within the trial was programmed as a random value within a uniform distribution of 0 to 50 days.
\item Toxicity Grade: computed for each patient on a daily basis for the extent of their assigned duration. To model  the trajectory of toxicity grade over time, we made the following simplifying assumptions:
\begin{enumerate}
\item On any given day, patients can rise or fall a single toxicity grade
\item Transitions in toxicity grade are random, but a larger hazard ratio suggests a greater chance of exacerbation and lower chance of recovery
\item A patient is censored once their pre-assigned duration within the trial has elapsed or they reach maximum toxicity, in this case representing death, whichever occurs first.
\end{enumerate}
\end{enumerate}

A hazard ratio for control:treatment is modeled for the control group to have higher toxicity burden through time compared to the treatment group (the value is programmed 1.0 or higher). For the control group, the probability of exacerbation is a base probability of 0.10 multiplied by the hazard ratio. Should exacerbation not occur, and the current stage is above the minimum, the probability of recovery is a base probability of 0.05 divided by the hazard ratio. Patients in the treatment group fluctuate based on base probabilities alone. Once a patient reaches the maximum toxicity or their maximum assigned duration, they are censored.

\subsection{Kaplan-Meier Estimator: Toxicity Trial}

We performed Kaplan-Meier estimation using the Python 3.7 library “lifelines”.\cite{lifelines} This library was used to plot survival probabilities and conduct logrank tests. Results were validated by assessing source code for accuracy and making a direct comparison to results from SPSS v26 (IBM Corp.).\cite{spss}

To permit comparison to KM, all patients begin the trial at stage 0, which represents grade 1 toxicity. An “event” was considered exacerbation to the next stage. Following event occurrence, patients were removed from analysis. Censoring is represented by a Wye-symbol ($\wye$).

\begin{figure}[h]
\centering
\includegraphics[scale=0.6]{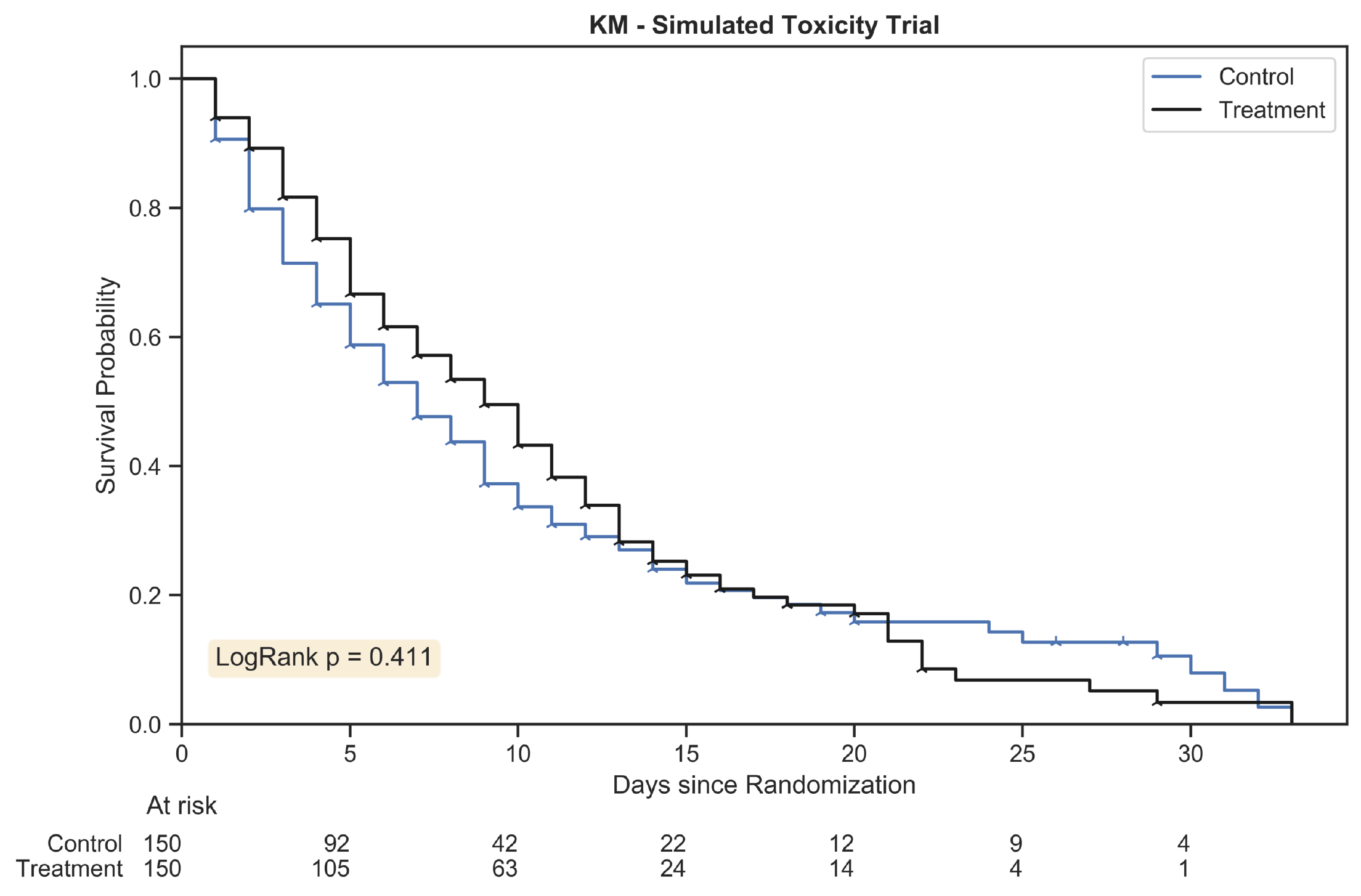}
\caption{The Kaplan Meier estimator plot for a randomly generated chemotherapy toxicity trial of 300 patients with 1:1 allocation. An event is considered the onset of chemotherapy toxicity (beyond stage 0) and patients are censored once their assigned duration has been reached. The hazard ratio between treatment arms is 1.25:1.\label{kmsim}}
\end{figure}

A single toxicity comparison trial was conducted with the following parameters: 200 patients (1:1 treatment allocation at 100 patients/arm) and a 1.25:1 hazard ratio for control:treatment. Fig. \ref{kmsim} depicts the corresponding Kaplan-Meier plot.

The outcome for a logrank test conducted on this trial is $P$ = 0.411; the result is not statistically significant. The Kaplan-Meier method is not sufficiently sensitive to distinguish between treatment arms for this simulated trial; high grades of toxicity may differ between the groups, but standard time-to-event statistics fail to capture the complex trajectory of morbidity.

Next, we analyze and report the identical drug trial using Weighted Trajectory Analysis.

\subsection{Weighted Trajectory Analysis: Simulated Trial}

\begin{figure}[h]
\centering
\includegraphics[scale=0.6]{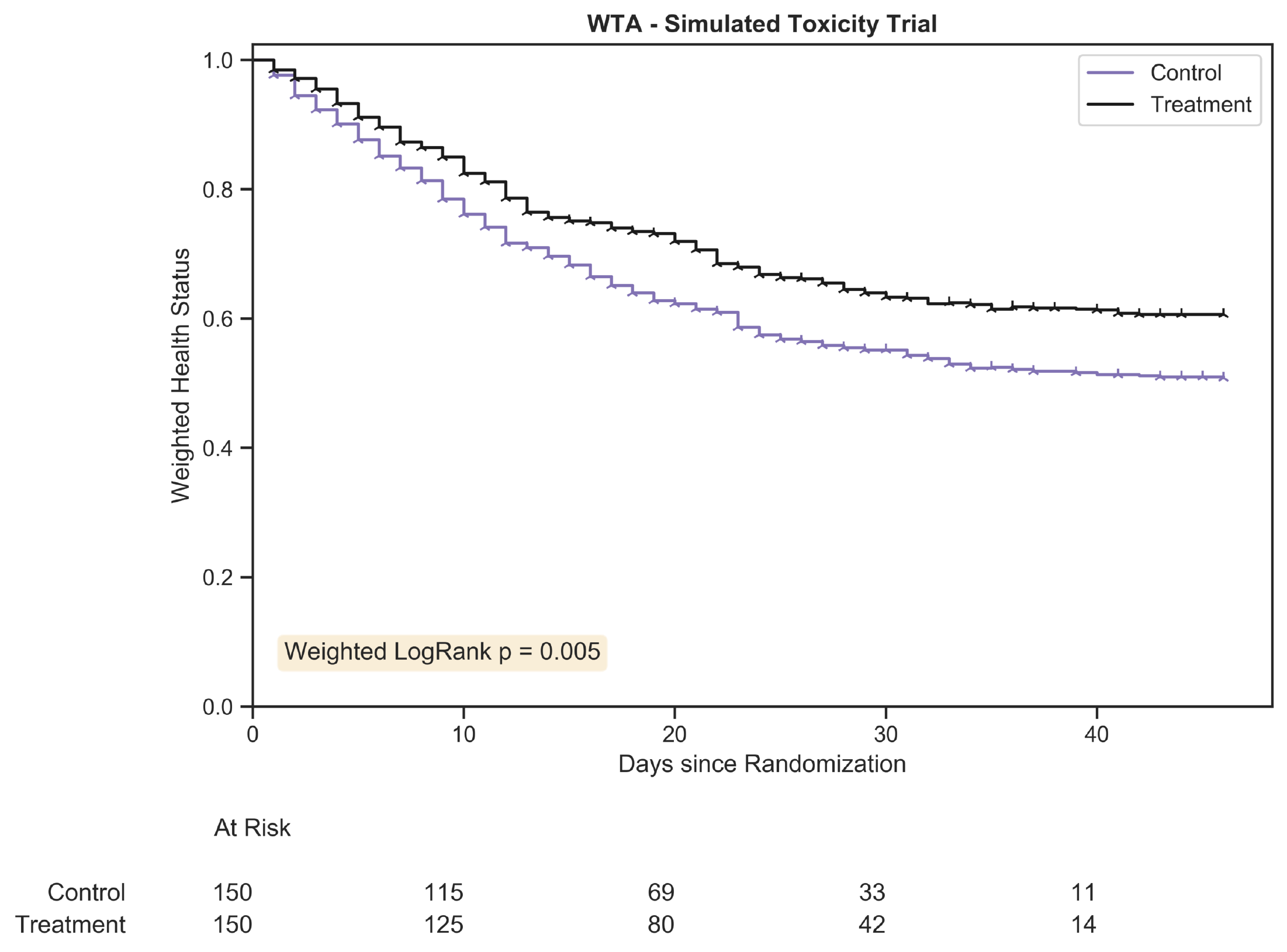}
\caption{The Weighted Trajectory Analysis plot for a randomly generated chemotherapy toxicity trial of 300 patients with 1:1 allocation. The weighted health status of both groups drop due to increasing morbidity from chemotherapy toxicity since randomization. The hazard ratio between treatment arms is 1.25:1. \label{wtasim}}
\end{figure}

The WTA is performed as described in the Section 2.3 on the identical trial dataset of 200 patients. Censoring is represented by a Wye-symbol ($\wye$) and occurs for each patient once they are no longer followed for toxicity grade. This takes places under two conditions: either the assigned duration for the patient has been reached, or the patient has suffered fatal toxicity. Fig. \ref{wtasim} provides the plot of WTA.

Note the change in x-axis range, number of patients at risk, and the trajectory of health status: patients are followed for the full course of toxicity and both declines and improvements are mapped. As compared to the KM plot, the treatment arms in this trial are visually distinct across all time points, demonstrating a reduced disease burden for the treatment group, a difference sustained across time. By approximately Day 30, a minor proportion of the original patients within the trial remain, and the delta between groups plateaus. Much like KM plot interpretation, the clinical significance of each trajectory drops after a substantial fraction of patients have been censored.

Using the “weighted” logrank test, $P$ = 0.005. WTA is a more powerful and more clinically relevant statistic for this dataset for its ability to track toxicity severity across all grades. As KM failed to reject the null hypothesis despite clinically meaningful group differences, a Type II error occurred. The improved sensitivity of WTA prevents such an error from taking place.

\subsection{1000-Fold Power Comparison - KM vs. WTA}

The trial analyzed in Sections 3.1 and 3.2 was a single instance of randomly generated data; the improved performance of WTA over KM may have occurred by chance. To accurately compare the ability of the tests to distinguish between treatment arms, we ran 1000-fold analyses across increments of sample size from 20 to 300 and hazard ratio from 1.0 to 1.5. For each trial, a p-value was computed using both KM and WTA. The fraction of tests that were significant (at $\alpha$ < 0.05) represents the power of the test (correctly rejecting the null hypothesis that the two groups are the same).

\begin{figure}[h]
\centering
\includegraphics[scale=0.6]{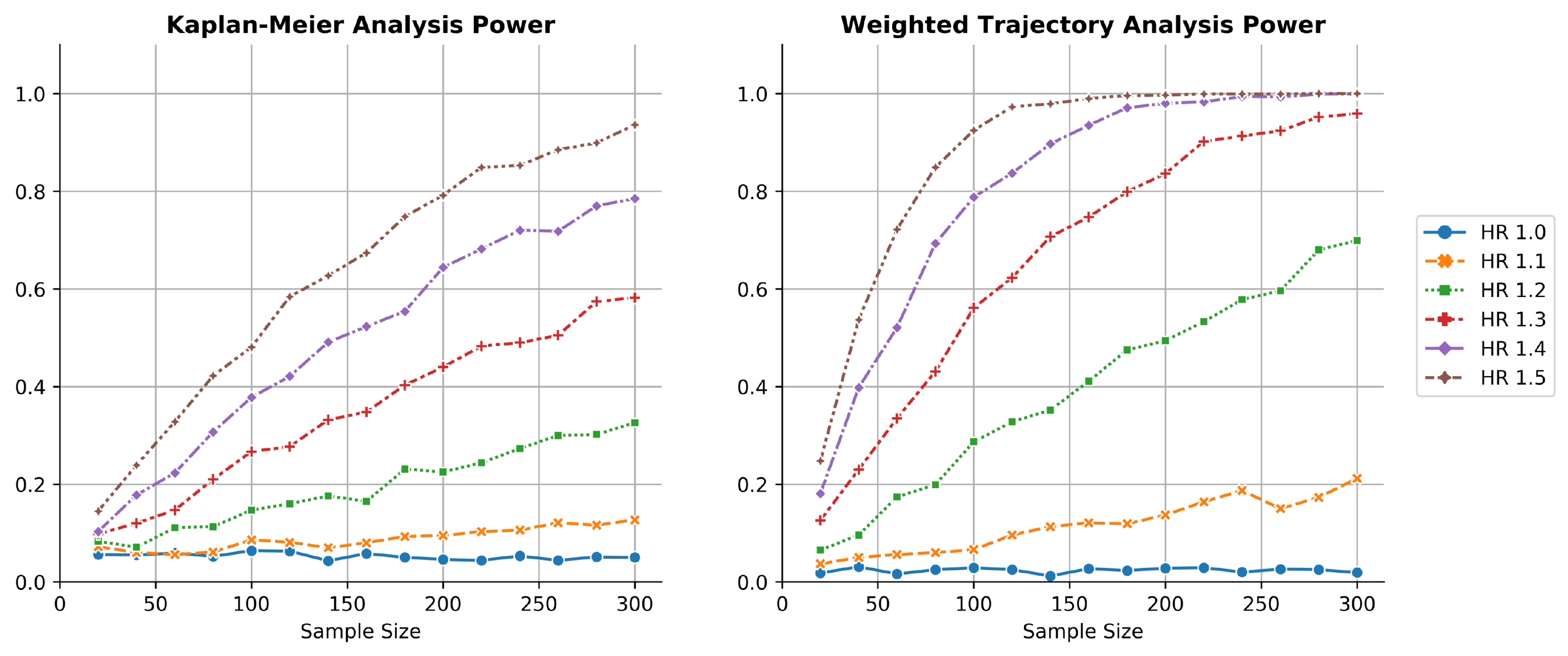}
\caption{1000-fold simulations of power as a function of sample size for both KM and WTA across several hazard ratios. WTA demonstrates consistently higher power, reflecting a smaller sample size requirement during trial design. The Type I error rate of WTA is approximately 0.025, indicating the method is conservative. The Type I error approaches 0.05 in the limit of larger trials with more distinct failure times.\label{powercomp}}
\end{figure}

Fig. \ref{powercomp} demonstrates that WTA has a consistently higher power than KM: it permits comparable analyses with a smaller sample size. Given that trial data is randomly generated, the plots are not perfectly smooth, but follow the expected logarithmic shape of power as a function of sample size.

For the simulated clinical trial at a 1.3 hazard ratio, WTA is able to reach 80$\%$ power at 180 patients while KM requires well over 300 patients. At a 1.4 hazard ratio, WTA requires about 100 patients for 80$\%$ power while KM requires about 300. Across many hazard ratios, WTA requires less than half the sample size to achieve a power equivalent to KM. Note that the power of the KM method for these clinical trials at a 1.5 hazard ratio mirrors the power of WTA at a 1.3 hazard ratio.

In this simulated example, Weighted Trajectory Analysis demonstrated greater sensitivity than Kaplan-Meier to a dataset with ordinal severity scoring. With a greater likelihood of correctly rejecting the null hypothesis, the novel method reduces Type II errors.

\subsection{1000-Fold Power Comparison - KM, WTA (Analytic and Computational), GEE}

To demonstrate the differences between the analytical and computational approach of WTA (and reference these against standard approaches of KM and GEE), we ran 1000-fold analyses under 9 unique conditions, at sample sizes of 100, 200, and 300 across hazard ratios of 1.0, 1.2, and 1.4. For each trial, a p-value was generated for all four of KM, WTA (analytical approach), WTA (simulated approach), and GEE longitudinal analysis using their respective hypothesis tests. The fraction of tests that were significant (at $\alpha$ < 0.05) represents the power of the test (correctly rejecting the null hypothesis that the two groups are the same).

\begin{figure}[h]
\centering
\includegraphics[scale=0.55]{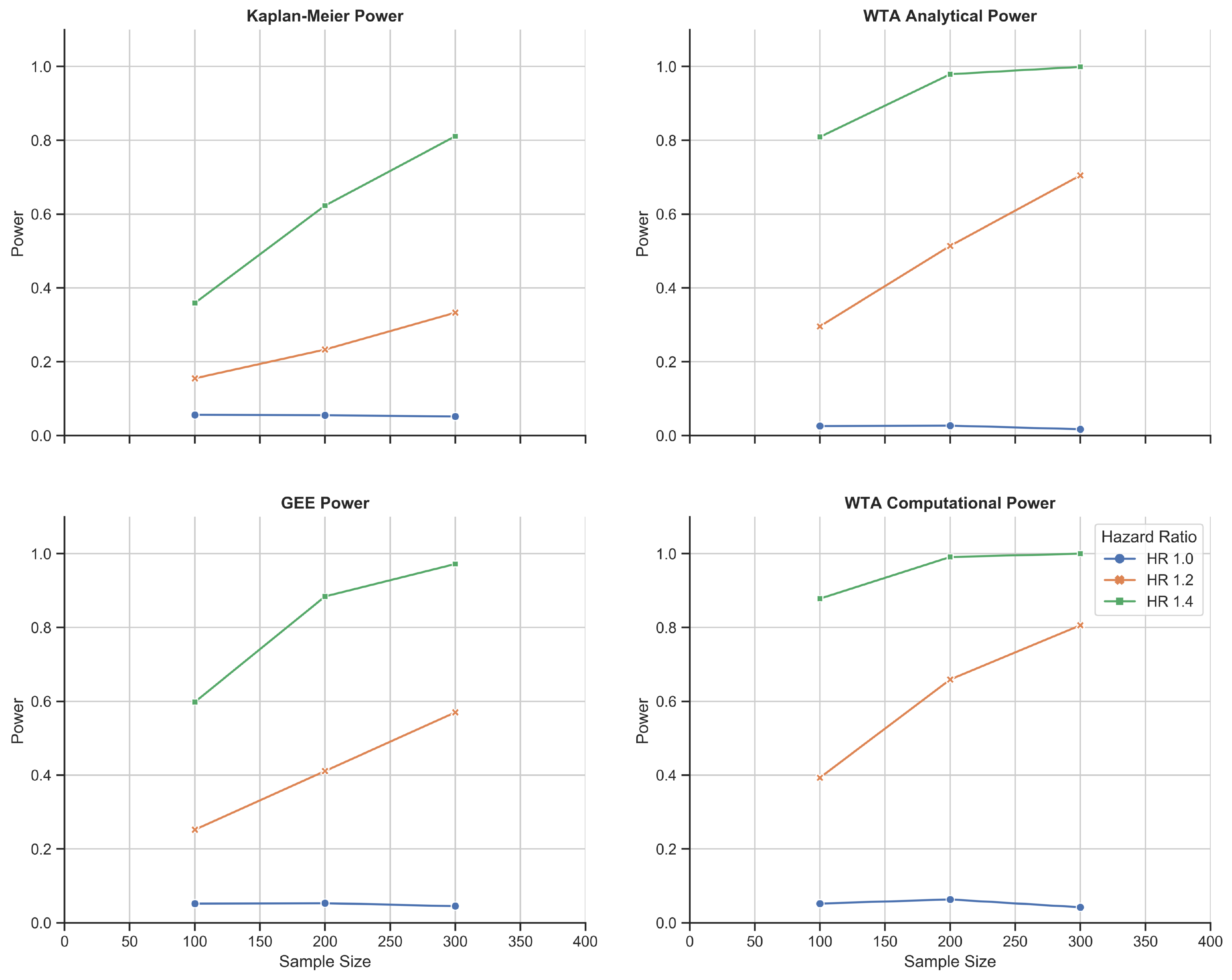}
\caption{Chemotherapy Toxicity Simulation Study: 1000-fold simulations of power as a function of sample size for KM, GEE, and WTA in both its analytical and computational form. WTA outperforms KM and GEE with consistently higher power and thus smaller sample size requirement. In addition, the computational approach of WTA outperforms the analytical approach in return for a more time and resource intensive methodology. The computational approach also meets a standard Type I error rate of 0.05 robust to changes in trial size.\label{toxkmwtagee}}
\end{figure}

Fig. \ref{toxkmwtagee} demonstrates that the analytical approach of WTA is less sensitive and less powerful than the computational approach. This is expected considering its computational effort and independence of trial size. Importantly, the analytical approach provides conservative results: in this stochastic model, the Type I error hovers around half of the 0.05 standard met by KM, GEE, and the computational approach of WTA. In the second simulation study, the explanation for this discrepancy becomes evident; the analytical approach is based on a normal approximation that becomes more precise with a larger number of distinct failure times and longer follow-up. As the second simulation study meets these criteria, the simulated Type I error correspondingly becomes closer to the 0.05 standard, the asymptotic limit.

GEE longitudinal analysis was found to be consistently weaker than both methods of WTA. This remains true in the second simulation study. The discrepancy is likely a trade-off on the parametric nature of each test: WTA is non-parametric and does not require any assumptions about survival outcomes. GEE is semi-parametric, which is less robust, but permits simultaneous modeling of multiple covariates as opposed to a sole comparison across treatment groups. As per this simulation study at a hazard ratio of 1.4, the analytical WTA meets the 80$\%$ power standard of clinical trial design at 100 patients; GEE requires over 150 patients and KM requires 300. The most accurate method, the computational WTA, requires fewer than 100 patients.

\section{Simulation Study 2 - Schizophrenia}\label{sec4}

The first simulation study highlighted the functionality of WTA under restrictive and common trial conditions to permit analysis with KM. However, some trials or datasets outside of medicine optimally analyzed using WTA may involve more extreme input parameters. Longer durations of patient participation are larger fluctuations within the data would also grant sensitivity to the analytical approach in Section 2.5. Accordingly, we developed a second simulation study to demonstrate the flexibility of WTA, in this case solely in analytic form, and compared its power to the versatile GEE longitudinal analysis.

The design is a Phase III comparison of antipsychotic efficacy in the management of schizophrenia. Compared to most chronic medical illnesses, psychiatric illness often demonstrates a more tumultuous course, with periods that may be completely asymptomatic interspersed with episodes of debilitating disease burden. Schizophrenia combines this generalization with a progressive disease course and often incomplete recovery following acute decompensations of the primary disorder or substance-induced episodes of psychosis.

As before, there are two treatment arms (control and treatment, 1:1 allocation). The variable of interest is symptom severity stage: stages range from 0 (absence of symptoms) to 6 (life-threatening illness due to severe disease burden and neurocognitive decline). Patients enter the trial at stage 2, which represents a symptom burden below full threshold for a psychotic episode; in our scenario, these patients are recruited for the trial due to a positive symptom screen as opposed to emergency psychiatric admission typical of greater symptom severity. Measurement intervals represent months as opposed to days, which permit larger transitions between stages in a single time interval, though loaded probabilities favour smaller transitions near extreme ends of the severity scale. Patients are enrolled into the trial for a randomized duration chosen from a uniform distribution between 36 and 84 months; they are censored when they reach the assigned duration or sooner if they reach stage 6. Mechanics of the study otherwise mirror Simulation Study 1.

\subsection{1000-Fold Power Comparison - WTA vs. GEE}

Once again, we ran 1000-fold analyses under 9 unique conditions, at sample sizes of 100, 200, and 300 across hazard ratios of 1.0, 1.2, and 1.4. For each trial, a p-value was generated for both WTA (analytical approach) and GEE longitudinal analysis using their respective hypothesis tests. The fraction of tests that were significant (at $\alpha$ < 0.05) represents the power of the test (correctly rejecting the null hypothesis that the two groups are the same).

\begin{figure}[h]
\centering
\includegraphics[scale=0.65]{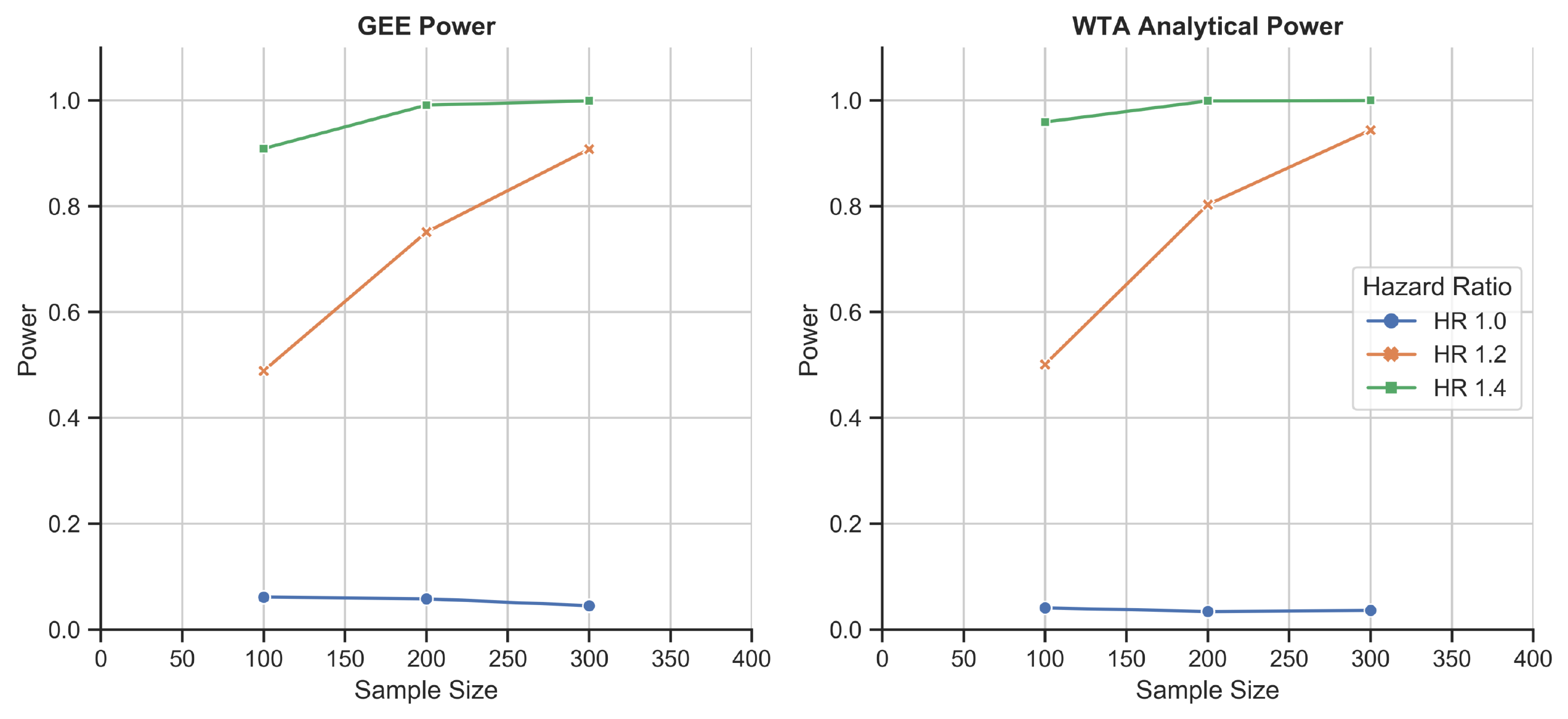}
\caption{Schizophrenia Disease Course Simulation Study: 1000-fold simulations of power as a function of sample size for GEE and WTA in its analytical form. WTA again outperforms GEE and demonstrates a Type I error rate of 0.037, closer to the 0.05 standard due to the larger size of each trial. \label{sczwtagee}}
\end{figure}

Fig. \ref{sczwtagee} demonstrates that under a vastly different stochastic model compared to the first simulation study, WTA once again outperforms GEE. The Type I error of WTA has shifted to an average of 0.037, closer to 0.05 given a trial with increased follow-up and failure times, which better satisfies the normal approximation underlying the method. This longer trial with more complex fluctuations in disease severity exhibits a higher power at identical hazard ratios and sample size compared to the previous study.

\section{Illustrative Real-World Example}\label{sec5}

\subsection{Immune Checkpoint Inhibitor Therapy for Melanoma}

Immune checkpoint inhibitors (ICIs) have transformed the treatment landscape for melanoma.\cite{jamaonc} Inhibitors targeting cytotoxic T lymphocyte antigen-4 (CTLA-4) and programmed death-1 (PD-1) produce a response in a large fraction of cancer patients. These responses are often durable and some are even curative. The use of Anti-CTLA-4 and Anti-PD-1 in combination has demonstrated the highest rate of durable response among melanoma treatment protocols. In prescribing a treatment plan, the promising response rates must be balanced with concerns about toxicity outcomes. Toxic effects associated with ICIs are immune-related in nature, may impact any organ, and remain a major challenge in clinical care.

Published data comparing therapy protocols suggests that the use of combination CTLA-4/PD-1 therapy results in significantly higher immune-related toxicity when compared to monotherapy regimens.\cite{larkin2015} These results may limit the use of combination therapy for patients with melanoma and remain a barrier to the development of new combinations. 

However, when treatment outcomes are compared over a longer time horizon, the discrepancy in immune-related toxicities seen between patients treated with combination versus monotherapy disappears. Those patients treated with combination therapy do experience greater toxicity during active treatment, but because the large majority of toxicities are reversible, the health status of patients treated with combination therapy improves with time. Longitudinally, patients treated with combination immunotherapy receive fewer actual treatment infusions, however treatment response rate is higher, and long-term survival comparable.\cite{larkin2019} Put simply, the combination of CTLA-4 and PD-1-directed immunotherapy has greater efficacy despite a significantly shorter duration of therapy, and despite an initial increase in immune-related toxicities the health status of patients who respond to therapy is excellent. The key limitation of existing statistical methods used to evaluate toxicity outcomes is the failure to capture improvement and accurately map changes through time.

\begin{figure}[h]
\centering
\includegraphics[scale=0.7]{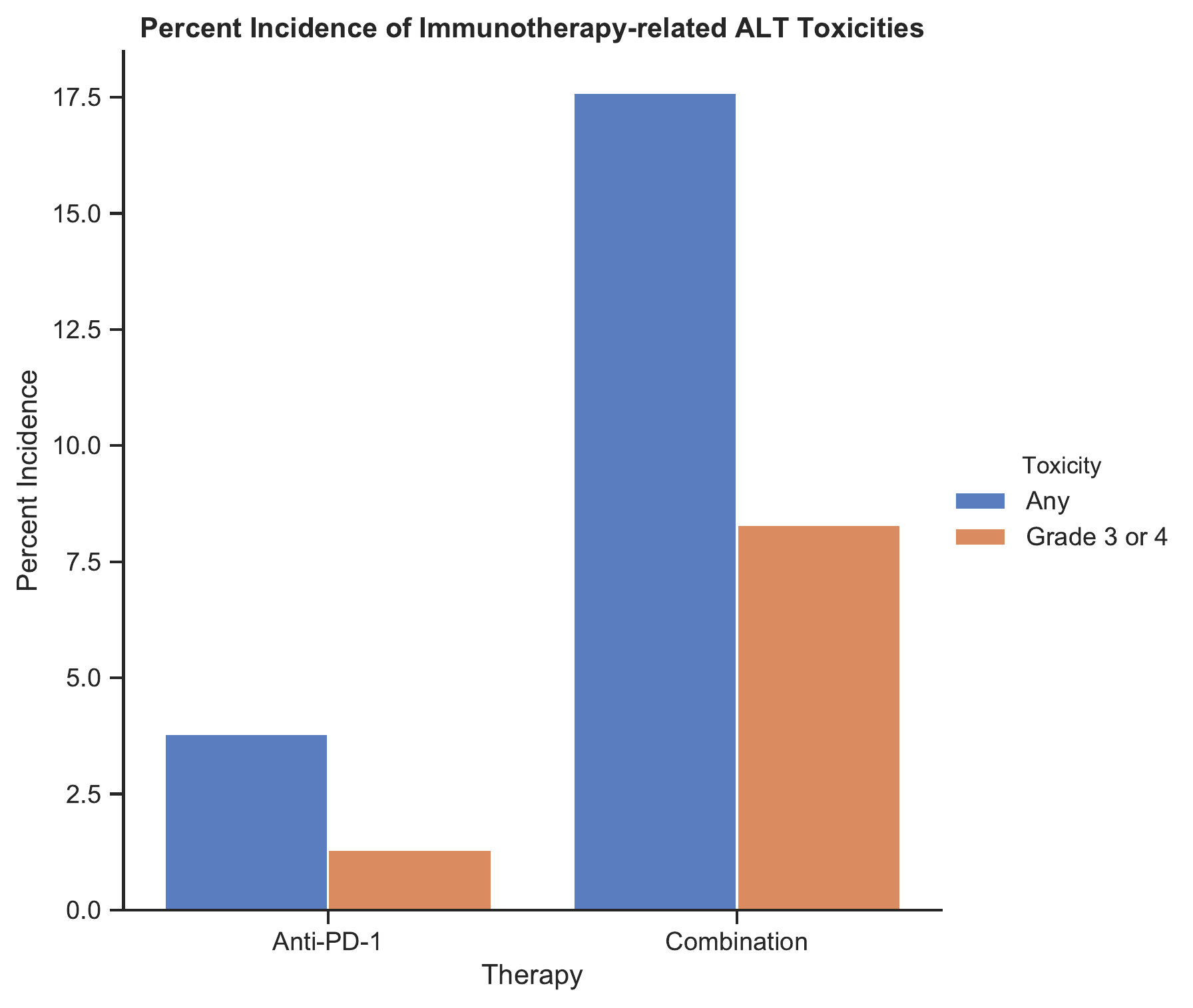}
\caption{The incidence of treatment-related toxicities associated with an increase in Alanine Aminotransferase (ALT) for patients receiving Anti-PD-1 therapy and combination therapy. Toxicities are graded using CTCAE v5.0.\cite{ctcae} Data from Table 3 of Larkin et al., 2015.\cite{larkin2015}\label{BarPlot_Melanoma}}
\end{figure}

The hypothesis that long-term health status is comparable between patients treated with combination versus monotherapy ICIs can be tested using Weighted Trajectory Analysis. Rather than using percent incidence to inform treatment decisions (see Figure \ref{BarPlot_Melanoma}), WTA will enable clinicians to assess the time-course of toxicity. The more detailed and sensitive mapping of toxicity outcomes will enable clinicians to more accurately translate patient data into standards for treatment.

In this example, retrospective toxicity data was used to compare monotherapy (Anti-PD-1) with combination therapy (Anti-PD-1 + Anti-CTLA-4). Increases in alanine aminotransferase (ALT) levels indicate transient, immune-related hepatitis, and were recorded for 195 melanoma patients on either protocol over 180 days. The increase in ALT from baseline was graded according to the National Cancer Institute Common Terminology Criteria for Adverse Events, version 5.0.\cite{ctcae} The baseline ALT scores are assigned a toxicity of 0 by definition. This enables comparison between KM and WTA.

\paragraph{Kaplan-Meier Estimator: Anti-PD-1 vs. Combination Therapy}

\begin{figure}[h]
\centering
\includegraphics[scale=0.8]{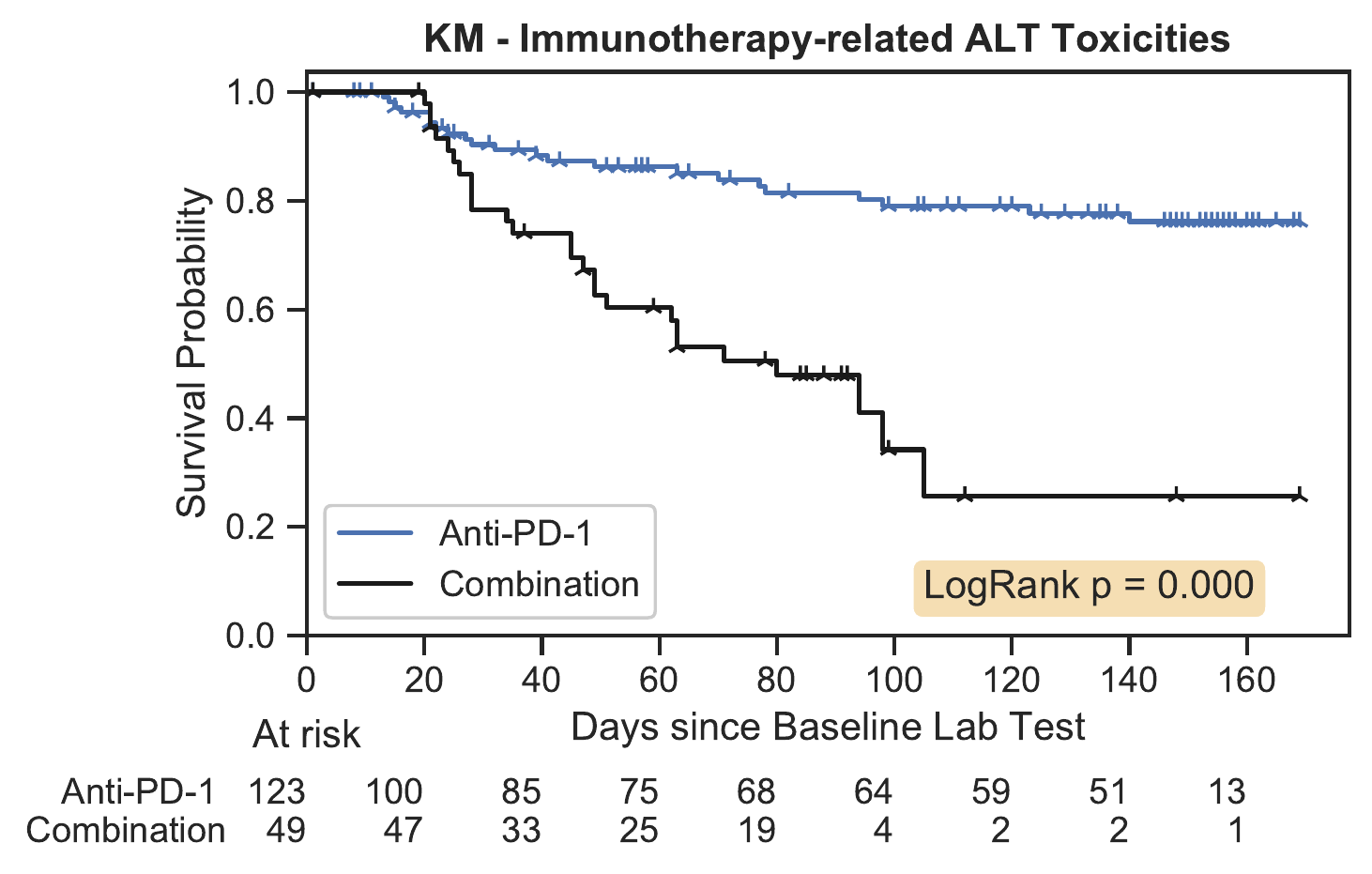}
\caption{The Kaplan Meier estimator plot for immunotherapy-related toxicities associated with an increase in ALT. An event is considered the onset of a nonzero toxicity grade.\label{KM_Melanoma}}
\end{figure}

To perform KM, the occurrence of any nonzero toxicity score was considered an event. The KM results in Figure \ref{KM_Melanoma} demonstrate that patients on combination therapy had a greater risk of experiencing nonzero toxicity over 100 days compared to the monotherapy group. This difference between groups was statistically significant with a p-value < 0.001.

\paragraph{Weighted Trajectory Analysis: Anti-PD-1 vs. Combination Therapy}

\begin{figure}[h]
\centering
\includegraphics[scale=0.7]{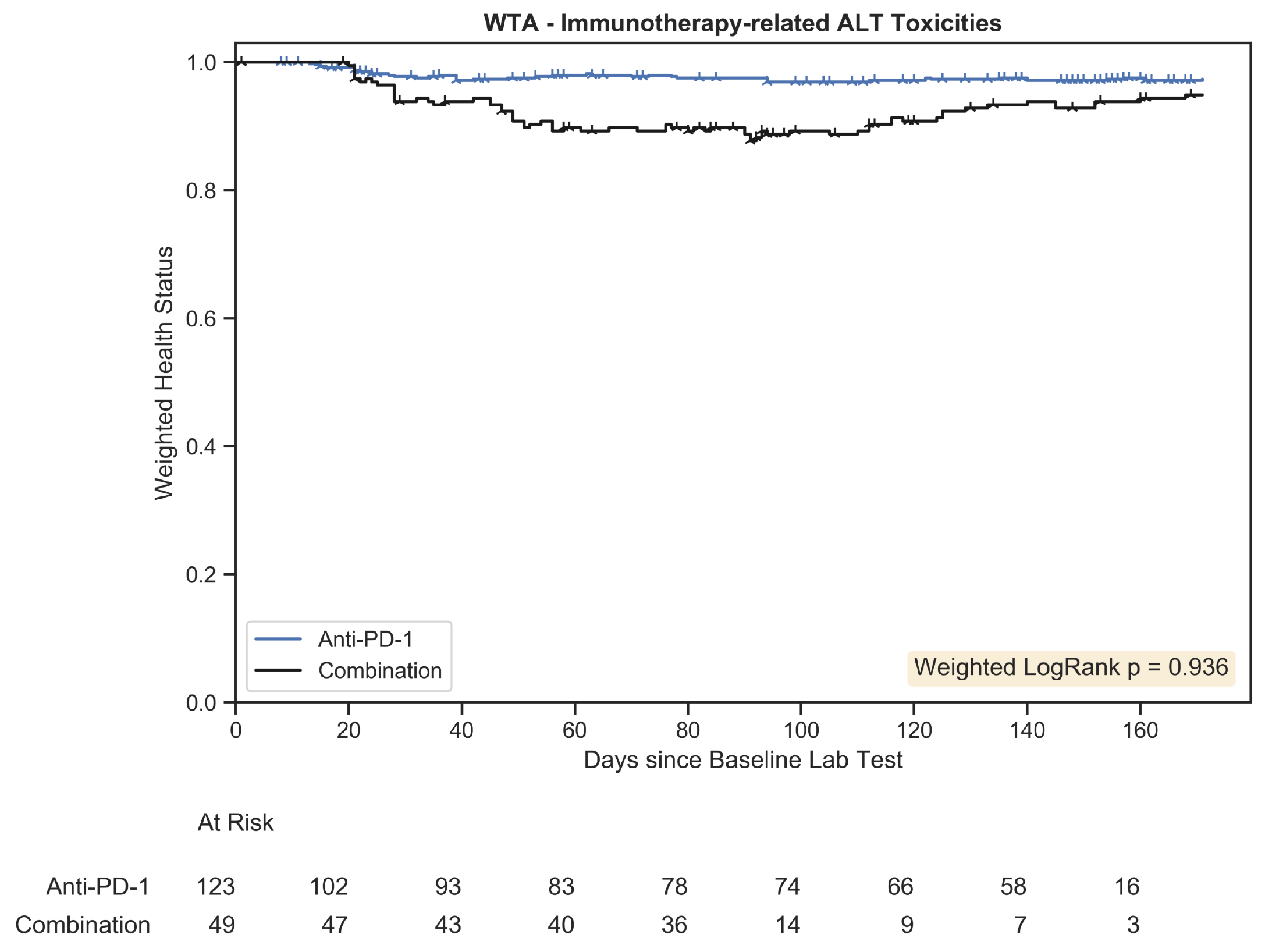}
\caption{Weighted Trajectory Analysis plot for immunotherapy-related toxicities associated with an increase in ALT. The weighted health status of the combination group initially diverges from the Anti-PD-1 group, but subsequent recovery leads to similar longitudinal outcomes.\label{WTA_Melanoma}}
\end{figure}

The WTA results are depicted in Figure \ref{WTA_Melanoma}. The Anti-PD-1 group has a steady accumulation of toxicity related events while the Combination group features a faster decline that plateaus at approximately 60 days. However, the trajectory of the Combination group recovers, and by 160 days, the two trajectories nearly converge. As immune-related toxicities are often reversible, the ability to model both exacerbation and recovery provides a more accurate picture of clinical outcomes.

The weighted logrank test had a p-value of 0.936, which is not statistically significant. The ability of recovery events to be captured within the weighted logrank hypothesis test demonstrates that differences in toxicity outcomes between these groups are misrepresented by prevalence data and the use of time-to-event curves like Kaplan-Meier. The absence of significant differences through more robust analysis suggests incidence data provides an incomplete picture of toxicity outcomes, leading to a false rejection of the null hypothesis. In the simulated example examining the development of toxicity to chemotherapy, WTA avoids a Type II error. In this real-world example, the use of WTA avoids a Type I error.

\subsection{ROSE/TRIO-012 Trial}

Treatment using agents that disrupt tumor angiogenesis (the process of generating new blood vessels) have shown clinical benefits in colorectal cancer, renal cell carcinoma, and several gynecological cancers.  The ROSE/TRIO-012 trial sought to evaluate ramucirumab, an anti-angiogenic drug, for the treatment of metastatic breast cancer.\cite{trio} Investigators compared ramucirumab to a placebo, when added to standard docetaxel chemotherapy. 

Many phase III trials within oncology are evaluated using Kaplan-Meier estimates and additional metrics based on the Response Evaluation Criteria in Solid Tumors (RECIST).\cite{recist} In ROSE/TRIO-012, KM was performed to determine progression free survival, in which disease progression and death are considered events, and overall survival, where death alone is an event. The RECIST framework (Table \ref{tab:recisttab}) was used to determine overall response metrics. These metrics reflect patients whose cancer improved through the course of the trial (objective response rate, ORR) and patients that did not experience progressive disease or death (disease control rate, DCR).

The ORR and DCR are defined as follows:

\begin{equation}
ORR = CR + PR
\label{orr}
\end{equation}
\begin{equation}
DCR = CR + PR + SD
\label{dcr}
\end{equation}
\\
Together, the several endpoints provide a detailed picture of patient outcomes since randomization. However, the individual metrics take time to interpret, and can sometimes provide conflicting signals regarding trial success. ROSE/TRIO-012 provides an example: although investigator-assessed PFS (p = 0.077) was insignificant at p < 0.05, endpoints including ORR and DCR were significantly higher in the ramucirumab group. 
The final verdict on the trial was that it had failed to meaningfully improve important clinical outcomes - a decision made solely on the absence of significance in investigator-assessed PFS, the trial's primary endpoint. Had trial success been defined as a composite of several endpoints, the investigators may have concluded that ramucirumab conferred a significant benefit to the patients within the study. Currently, ramucirumab is not approved for use in the treatment of metastatic breast cancer.

The ability to combine the RECIST framework with mortality in a single plot would allow oncologists to rapidly interpret the totality of results of a clinical trial. A judgment on trial success can remain tied to the significance of a primary objective, but this objective should capture a wide array of important patient outcomes. In this example, ROSE/TRIO-012 trial results from Mackey et al's 2014 paper are compared to Weighted Trajectory Analysis on the original data.

\paragraph{Kaplan-Meier: Ramucirumab vs. Placebo + Docetaxel}

Figures 2A and 2C of Mackey et al.'s 2014 paper are provided below. Respectively, they represent progression-free survival (the primary endpoint) and overall survival, both using standard Kaplan-Meier techniques. Upon inspection, progression free survival appears slightly higher within the ramucirumab group. The logrank p-value of 0.077 did not indicate statistical significance. As PFS was the primary endpoint, the intervention was deemed unsuccessful. Overall survival outcomes were no different between groups (p = 0.915).

\paragraph{RECIST Endpoints: Ramucirumab vs. Placebo + Docetaxel}

Conflicting signals about the efficacy of ramucirumab arise when analyzing secondary endpoints. ORR and DCR were significantly higher in the ramucirumab arm (44.7\% vs. 37.9\%, p = 0.027; 86.4\% vs. 81.3\%, p = 0.022).

ORR and DCR provide no time-to-event information. The goal of combining RECIST metrics with KM is to generate a complete picture of patient outcomes. However, by omitting information on time and severity, respectively, the distinct methods may disagree on intervention efficacy. The whole is less than the sum of its parts. 

The existing solution to this apparent conflict is a decision made by the investigators prior to the study: select a single metric as the primary objective to determine success. This both focuses and simplifies any conversation about study outcomes. Had this primary objective been ORR, the conclusion of the study would have supported the use of ramucirumab for these patients.

\paragraph{Weighted Trajectory Analysis: Ramucirumab vs. Placebo in addition to Docetaxel}

We used Weighted Trajectory Analysis to combine the RECIST framework with mortality to depict comprehensive time-to-event outcomes. To perform the method, we assign the following ordinal severity scoring framework:

\begin{table}[h]
\centering
\begin{tabular}{l|l}
\textbf{Outcome} & \textbf{Score}  \\ 
\hline
CR (Complete Response)               & 0                       \\
PR (Partial Response)               & 1                       \\
SD (Stable Disease)              & 2                       \\
PD (Progressive Disease)               & 3                       \\
Death            & 4                      
\end{tabular}
\end{table}

The starting point of each patient at the time of randomization is stable disease (SD), a score of 2. At the ends of the ordinal scale are complete response (CR, the best outcome) and death (the worst outcome). Patients are censored upon withdrawal, loss to follow-up, or directly following death.

Using the original ROSE/TRIO-012 dataset and Table \ref{comptab}, we generate Figure \ref{WTA_Trio}. Censoring is indicated using vertical tick marks.

This plot provides a comprehensive view of all patient outcomes for the full study duration. A few months into the trial we see the peak in weighted health status for both groups. This occurs at 68 days for the placebo group and 76 days for the ramucirumab group. At this phase, some patients have experienced partial or complete response. Following this peak is a gradual descent that represents progressively increasing morbidity and death across both groups. The trajectories are strikingly similar, with the ramucirumab group experiencing slightly better outcomes throughout the study. The difference is not statistically significant (p = 0.587). This corroborates the current regulatory standard that ramucirumab should not be approved for the treatment of metastatic breast cancer.

With the WTA plot alone, investigators can easily interpret the time course of disease response. Patients likely to respond or recover generally do so following the first two chemotherapy cycles. After three months, the prognosis is poor: both treatment arms are characterized by progressive disease and death.

\section{Discussion}\label{sec6}

WTA was created to (a) evaluate Phase III clinical trials that assess outcomes defined by various ordinal grades (or stages) of severity; (b) permit continued analysis of participants following changes in the variable of interest; (c) demonstrate the ability of an intervention to both prevent the exacerbation of outcomes and improve recovery and the time course of these effects. Its development was inspired by a pressure injury study -- a disease process characterized by several stages of severity -- for which Kaplan-Meier estimates would fail to capture complete trajectory. Despite its limitations, KM provides crucial advantages such as patient censoring, rapid interpretation of a survival plot, and a simple hypothesis test. To this end, we sought to create a statistical method that built on the foundations of Kaplan-Meier analysis, but would overcome the inherent limitations of the technique.

We built the WTA toolkit based on expansion and extension of the Kaplan-Meier methodology.  We adapted the KM to support analysis of ordinal variables by redefining events as a change in disease score rather than assigning "1" and omitting the patient from further analysis. We adapted the KM to permit fluctuating outcomes (worsening and improvement of the ordinal outcome) by plotting a novel weighted health status as opposed to probability.  We retained the ability to censor patients at the time of non-informative status. These changes warranted a novel significance test, for which we developed a modification to Peto et al.'s logrank test.\cite{peto2} This analytical approach is rather conservative in its Type I error rates for smaller trials, but the rate approaches 0.05 in the limit of massive trials with many distinct failure times. Thus, we developed a computational approach that is more resource intensive but remains precise and accurate independent of trial size.

In order to explore and demonstrate the utility of WTA, we applied WTA to two randomized clinical trial simulation studies. The first clinical setting was chemotherapy toxicity, a trial in which the variable of interest ranged from 1-5 (shifted to 0-4), stage transitions were singular and started at 0, and up to 50 discrete time points were measured for each patient. The second setting was schizophrenia disease course, a more complex trial in which the variable of interested ranged from 0-6, stage transitions were often multiple and started at 2, and up to 84 discrete time points were measured for each patient. We performed sensitivity and power comparisons across both sample size and hazard ratio. Through 1000-fold validation, WTA showed greater sensitivity and power, often requiring fewer than half the patients for comparable power to KM. WTA also showed increased power compared to GEE, likely secondary to its more robust non-parametric methodology compared to the semi-parametric GEE, at the cost of GEE’s ability to model covariate effects. This demonstrates that designing a Phase III clinical trial using our novel method as the primary endpoint can substantially lower cost, duration, and the risk of Type II errors.

We also applied WTA to real-world clinical trial data. The first was the assessment of time-dependent toxicity grades in melanoma patients receiving one of two immunotherapy treatment regimens. Although toxicities are generally reported in oncology trials as the worst grade experienced by each individual patient, this fails to capture those toxicities that resolve with treatment modification or targeted intervention. As such, the published literature suggested prohibitive toxicity of the most effective therapy, while practitioners’ experience was that high grade toxicities were often transient and treatable. The WTA we conducted confirmed that treatment-related toxicities of combination therapy resolved to rates close to that seen with less effective monotherapy regimens. The second was the re-evaluation of a published phase III registration trial or an antiangiogenic drug for the treatment of metastatic breast cancer. Although this study failure to demonstrate statistically significant improvement in the pre-defined primary endpoint, a number of secondary endpoints suggested the possibility of meaningful clinical benefit from the antiangiogenic therapy. By using an ordinal scale to describe the spectrum of clinical outcomes after therapy, spanning complete disease response, partial response, disease stability, disease progression, and death, WTA demonstrated that although patients derived a modest benefit from antiangiogenic therapy when compared to control therapy, the difference was neither clinically nor statistically significant. The resulting graph captures the full clinical course of patients in a single figure. This result underscores that WTA did not inappropriately provide an overly sensitive analytic tool and justified the regulatory stance that the intervention did not warrant approval to market. Overall, the novel method affords greater specificity and reduces the likelihood of Type I errors.

In aggregate, we feel the strengths of the Weighted Trajectory Analysis statistic are its ability to capture detailed trajectory outcomes in a simple summary plot, its greater power, and its ability to map exacerbation and improvement. These strengths are built upon key advantages that make KM a favored tool for clinical trial evaluation: namely, the ability to censor patients and a compare treatment arms using a simple hypothesis test. WTA-dependent trial design can substantially reduce sample size requirements, raising the practicality and lowering the cost of Phase III clinical trials. However, we acknowledge several limitations to this method. WTA does not facilitate Cox regression analysis or generate the equivalent of a hazard ratio. The WTA is a new technique and does not yet have a clinical or regulatory track record. WTA relies on the assumption of non-informative censoring, and investigation into alternative approaches to censoring such as inverse-probability of censoring weighting (IPCW) remains important future work.\cite{ipcw} Lastly, the WTA requires an assumption that the change between adjacent ordinal severities is equally important independent of the levels transitioned by applying a direct numerical weight. This conversation is not always medically appropriate: taking the example of pressure injuries, a transition from Stage 0 to 1 may necessitate a topical ointment, whereas a transition from Stage 3 to 4 warrants surgical repair. Thus, the method relies on a simplifying assumption and future research will be conducted to evaluate non-linear scoring systems. For multi-stage systems, this method remains more precise than collapsing scores to binary systems in order to use KM. Alternative statistical methods such as multi-state modelling are recommended to elicit transition intensities of each unique level as necessary. To encourage the evaluation and improvement of WTA, software is in development permit biostatisticians to further test, apply, and potentially expand the utility of WTA.

In summary, we report the development and validation of a flexible new analytic tool for analysis of clinical datasets that permits high sensitivity assessment of ordinal time dependent outcomes. We see multiple clinical applications, and have successfully applied the new tool in the analysis of both simulated and real-world studies with complex illness trajectories. Future direction with Weighted Trajectory Analysis includes the addition of confidence intervals to group trajectories, non-linear weights to mirror disease burden, exploration of alternative censoring assumptions, and a regression method analogous to the Cox model.

\section*{Acknowledgments}

The authors thank Dr. David Vock (University of Minnesota), Drs. Edward Mascha and Chase Donaldson (Cleveland Clinic), and Dr. Sunita Ghosh (University of Alberta) for helpful early discussions.

The authors also thank Britsol Myers Squibb for access to their melanoma clinical trial dataset and the TRIO-012/ROSE study team along with the TRIO Science Committee for access to their database.

\subsection*{Conflict of interest}

None declared.

\subsection*{Data Availability Statement}

The data that support the findings of this research are available on request from the corresponding author. The data are not publicly available due to privacy or ethical restrictions.

\nocite{*}
\bibliography{wileyNJD-AMA}

\newpage

\section{Tables}\label{sec7}

\begin{table}[h]
\caption{Feature comparison between the Kaplan-Meier Estimator and Weighted Trajectory Analysis.\label{comptab}}
\begin{tabular}{|l|l|l|}
\hline
\textbf{Feature}      & \textbf{Kaplan-Meier Estimator}                                                                                                                           & \textbf{Weighted Trajectory Analysis}                                                                                                                                                                        \\ \hline
Event                 & \begin{tabular}[c]{@{}l@{}}Outcome with binary coding. A patient must\\ begin at “0” and is removed from analysis\\ following an event (“1”).\end{tabular} & \begin{tabular}[c]{@{}l@{}}An event is a change in clinical severity and\\ does not remove a patient from further analysis. \\ Must be discrete with a finite range\\ that depends on the variable of interest.\end{tabular} \\ \hline
Variable of Interest  & \begin{tabular}[c]{@{}l@{}}Death, metastases, local recurrence,\\ stroke, and more. Can include variables\\ outside of medicine, such as post-graduate\\ employment.\end{tabular}                                    & \begin{tabular}[c]{@{}l@{}}Graded/staged outcomes: \\ ECOG performance, Toxicities, NYHA Heart\\ Failure Class, Questionnaire scores, and more;\\ also includes variables outside of medicine.\end{tabular}                                                               \\ \hline
Trajectory            & Survival function always decreases.                                                                                                              & \begin{tabular}[c]{@{}l@{}}Bidirectional: severity function\\ can decrease or increase.\end{tabular}                                                                                                          \\ \hline
Censoring             & \multicolumn{2}{l|}{Removes patients from subsequent analysis (for withdrawal, discharge, lost to follow-up, etc.).}                                                                                                                                                                                                                                            \\ \hline
Test for Significance & Logrank Test                                                                                                                                    & Weighted Logrank Test                                                                                                                                                                                        \\ \hline
Y-axis                & Survival Probability                                                                                                                            & Weighted Health Status                                                                                                                                                                                         \\ \hline
X-axis                & \multicolumn{2}{l|}{Time (discrete: days, weeks, months, etc.)}                                                                                                                                                                                                                                                                                                                                      \\ \hline
Y-intercept           & 1.0                                                                                                                                             & Between 0 and 1.0, inclusive \\ \hline
\end{tabular}
\end{table}

\newpage

\begin{table}[htbp]
  \centering
  \caption{A snapshot of the final results of a simulated chemotherapy toxicity grade trial. }
    \begin{tabular}{|c|c|c|c|c|c|c|c|c|c|c|c|c|c|}
    \toprule
    \multicolumn{1}{|p{4.5em}|}{\textbf{Patient ID}} & \multicolumn{1}{p{7em}|}{\textbf{Treatment Arm}} & \multicolumn{1}{p{4em}|}{\textbf{Duration}} & \textbf{0} & \textbf{1} & \textbf{2} & \textbf{3} & \textbf{4} & \textbf{5} & \textbf{6} & \textbf{7} & \textbf{8} & \textbf{9} & \textbf{10} \\
    \midrule
    1     & 1     & 10    & 0     & 0     & 0     & 0     & 0     & 0     & 1     & 1     & 1     & 0     &  \\
    2     & 1     & 10    & 0     & 0     & 0     & 0     & 0     & 1     & 1     & 1     & 1     & 1     &  \\
    3     & 0     & 11    & 0     & 0     & 0     & 0     & 0     & 0     & 0     & 0     & 0     & 0     & 0 \\
    4     & 1     & 6     & 0     & 0     & 0     & 0     & 0     & 0     &       &       &       &       &  \\
    5     & 0     & 13    & 0     & 0     & 0     & 0     & 0     & 0     & 0     & 0     & 0     & 1     & 1 \\
    6     & 1     & 9     & 0     & 0     & 0     & 0     & 0     & 0     & 0     & 0     & 0     &       &  \\
    7     & 0     & 18    & 0     & 0     & 0     & 0     & 0     & 0     & 1     & 1     & 1     & 2     & 2 \\
    8     & 1     & 6     & 0     & 0     & 0     & 0     & 0     & 0     &       &       &       &       &  \\
    9     & 1     & 29    & 0     & 0     & 0     & 0     & 0     & 0     & 0     & 0     & 0     & 1     & 0 \\
    10    & 0    & 4     & 0     & 0     & 0     & 0     &       &       &       &       &       &       &  \\
    \bottomrule
    \end{tabular}%
  \label{snapshot}%
  \caption*{\footnotesize{Treatment Arms 0 and 1 represent the control and treatment groups, respectively. Numbered columns indicate sequential days within the trial starting at Day 0. Duration indicates the number of days the patient was hospitalized.}}
\end{table}%

\newpage

\begin{table}[]
\caption{RECIST 1.1 criteria definitions.}
\label{tab:recisttab}
\begin{tabular}{@{}ll@{}}
\toprule
\textbf{Treatment Outcome} &
  \textbf{Definition} \\ \midrule
Complete Response (CR) &
  \begin{tabular}[c]{@{}l@{}}Disappearance of all target lesions. Any pathological lymph nodes (whether target or non-target)\\  must have reduction in short axis to \textless{}10 mm.\end{tabular} \\ \midrule
Partial Response (PR) &
  \begin{tabular}[c]{@{}l@{}}At least a 30\% decrease in the sum of diameters of target lesions, taking as reference the baseline\\ sum diameters.\end{tabular} \\ \midrule
Progressive Disease (PD) &
  \begin{tabular}[c]{@{}l@{}}At least a 20\% increase in the sum of diameters of target lesions, taking as reference the smallest\\ sum on study (this includes the baseline sum that is the smallest on study). In addition to the\\ relative increase of 20\%, the sum must also demonstrate an absolute increase of at least 5 mm.\\ (Note: The appearance of 1 or more new lesions is also considered progression).\end{tabular} \\ \midrule
Stable Disease (SD) &
  \begin{tabular}[c]{@{}l@{}}Neither sufficient shrinkage to qualify for PR nor sufficient increase to qualify for PD, taking as\\ reference the smallest sum of diameters while on study.\end{tabular} \\ \bottomrule
\end{tabular}
\end{table}

\clearpage

\section{Figures}\label{sec8}

\clearpage

\clearpage

\clearpage

\clearpage

\clearpage

\clearpage

\clearpage

\clearpage

\begin{figure}[h]
\centering
\includegraphics[scale=1]{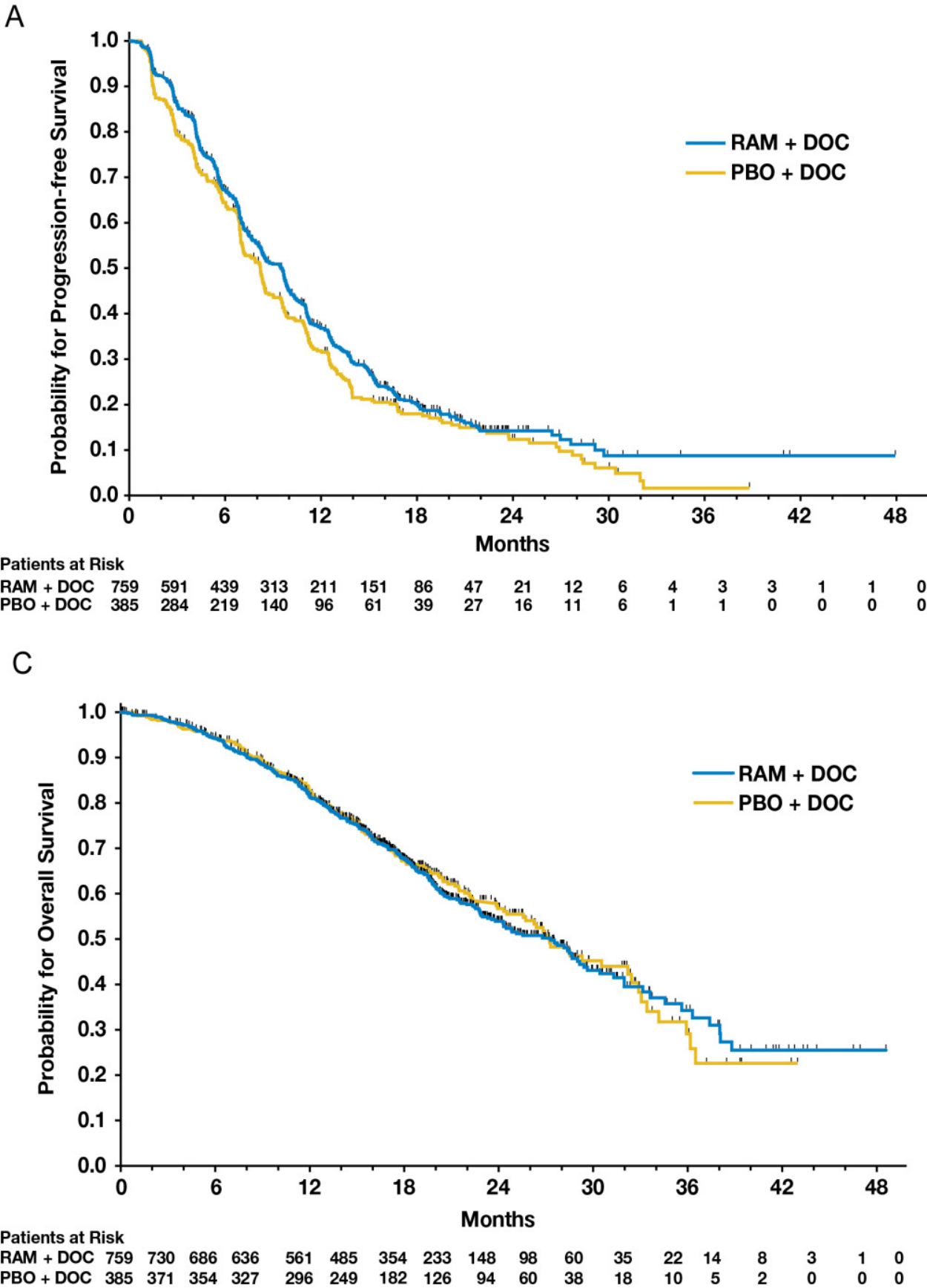}
\caption{Figures 2A and 2C of Mackey et al.'s 2014 paper comparing ramucirumab to a placebo added to standard docetaxel chemotherapy.\cite{trio} The figures provide patient outcomes using KM estimates of progression free survival (PFS) and overall survival (OS), respectively. \label{ramu}}
\end{figure}

\clearpage

\begin{figure}[h]
\centering
\includegraphics[scale=0.7]{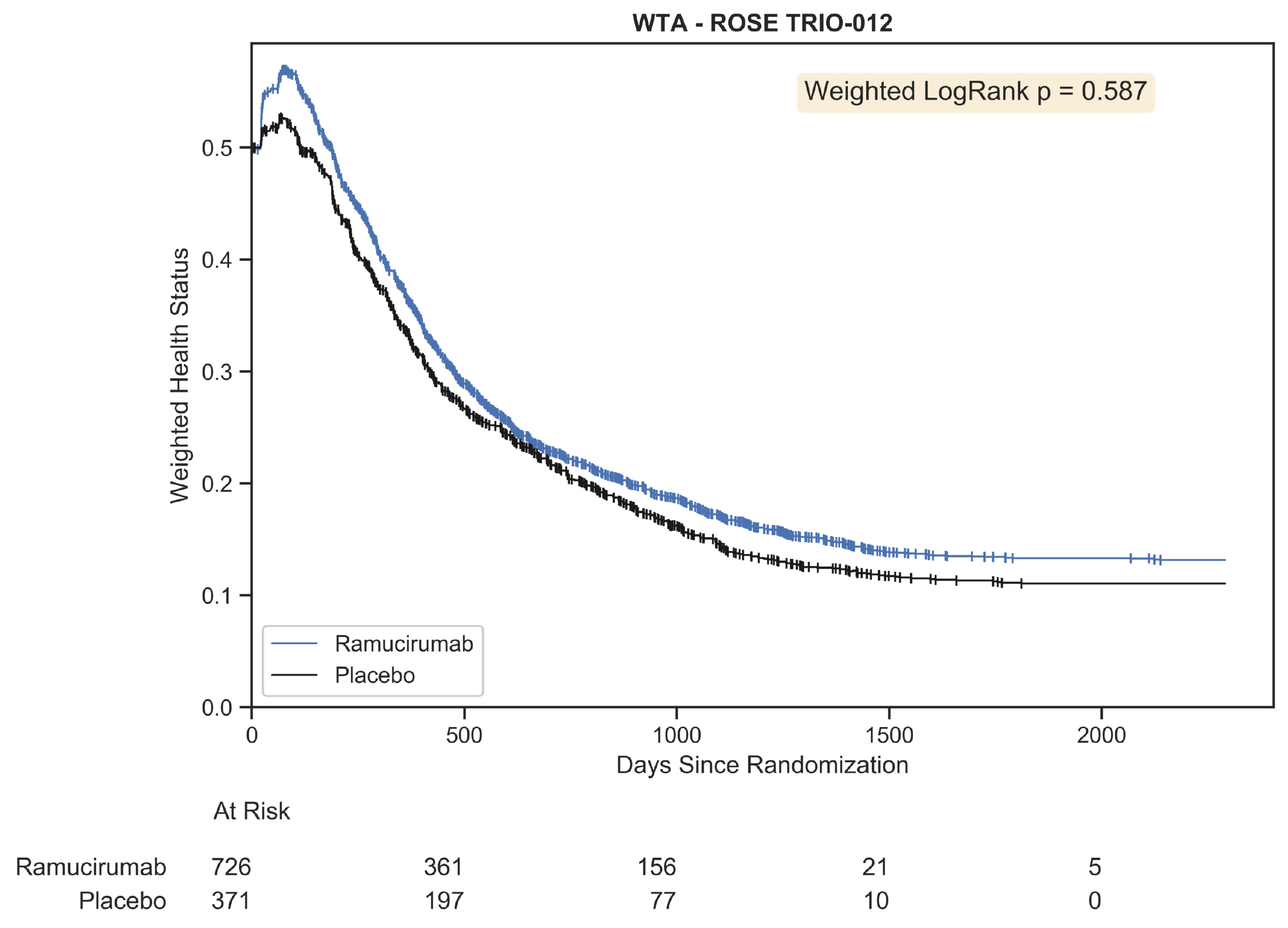}
\caption{Weighted Trajectory Analysis on the original ROSE/TRIO-012 dataset using an ordinal scale that merges RECIST criteria with mortality. The trajectory of patient outcomes demonstrates that partial and complete response initially outweigh progressive disease and mortality for the first few chemotherapy cycles. Following this peak, patient prognosis is generally poor as both treatment arms experience growing disease burden and death.\label{WTA_Trio}}
\end{figure}

\end{document}